\documentstyle[12pt]{article}
\def\c#1{\setbox0=\hbox{#1}\ifdim\ht0=1ex \accent'30 #1\else{\ooalign{\hidewidth\char'30\hidewidth\crcr\unhbox0}}\fi}

\input{tcilatex}

\begin{document}

\title{Spatial behaviour in dynamical thermoelasticity backward in time for porous
media \thanks{%
Work performed in the context of the COFIN MIUR, Italy, "Mathematical Models
for Materials Science" and under auspices of G. N. F. M. of the Italian
Research Council (C. N. R.).}}
\author{Gerardo Iovane, \quad Francesca Passarella \thanks{
E-mail address: iovane@diima.unisa.it, passarella@diima.unisa.it}}
\date{{\small Department of Information Engineering and Applied Mathematics
(DIIMA),} \\
{\small University of Salerno, 84084 Fisciano (Sa) Italy} }
\maketitle

\begin{quote}
{\bf Abstract} -- The aim of this paper is to study the spatial behaviour of
the solutions to the boundary--final value problems associated with the
linear theory of elastic materials with voids. More precisely the present
study is devoted to porous materials with a memory effect for the intrinsic
equilibrated body forces. An appropriate time--weighted volume measure is
associated with the backward in time thermoelastic processes.

Then, a first--order partial differential inequality in terms of such
measure is established and further is shown how it implies the spatial
exponential decay of the thermoelastic process in question.
\end{quote}

\section{Introduction}

The boundary--final value problems associated with the linear
thermoelasticity have been studied by Ames and Payne \cite{A91} in
connection with the continuous dependence of the thermoelastic processes
backward in time with respect to the final data. It is well known that this
is an improperly posed problem. A further study on this subject was recently
developed by Ciarletta \cite{ciarletta1}.

More recently, Chirita and Ciarletta have developed an exhaustive
description for the spatial behaviour of solutions in linear
thermoelasticity forward in time \cite{C99a,C99b,C99c}. In this connection
some appropriate time--weighted surface power functions are introduced. On
this basis some spatial decay estimates of Saint--Venant type are
established for bounded bodies; while for unbounded bodies some alternatives
of Phragm\'{e}n--Lindel\"{o}f type are obtained. In the present paper, we
aim to extend the above context to porous materials for backward in time
processes. We shall refer to the well-known theory by Nunziato and Cowin, in
which the presence of small pores (or voids) in the conventional continuum
model is introduced by assigning an additional degree of freedom, namely,
the fraction of elementary volume that is possibly found void of matter \cite
{[7],[8],[9]}. Following a previous paper by Goodman and Cowin on granular
(flowing) materials, the bulk mass density of the material is represented as
a product of two fields, the void volume fraction and the mass density of
the matrix material \cite{[7]}. Starting from the fundamental work of Iesan
in \cite{Iesan}, we consider the porous materials with the memory effects
for the intrinsic equilibrated body forces and we associate with such a
model the boundary-final value problem as in \cite{[9]} and \cite
{IesanCiarletta}. Then, we study the spatial behaviour of the thermoelastic
processes backward in time by using an appropriate time-weighted measure.

In Section 2, we present the basic equations of the linear dynamic theory of
porous body developed in \cite{[9]} and \cite{IesanCiarletta}. Some
constitutive assumptions and other useful results are also presented. In
Section 3 a time--weighted volume measure is defined and a first--order
partial differential inequality is established in terms of such measure. On
this basis a spatial decay estimate of Saint--Venant type is proved; our
decay results look quite similar to those established by Lin and Payne for
the backward heat equation defined on a semi--infinite cylinder \cite{LinPay}%
.

\bigskip

\section{Preliminaries}

We shall denote by $B$ the smooth domain of the physical space $(\equiv {\rm %
I\!R}^{3})$ occupied by an anisotropic, homogeneous and porous body in a
fixed, natural reference configuration. Identified ${\rm I\!R}^{3}$ with the
associated vector space, an orthonormal frame of reference is introduced.
The vectors and tensors will have components denoted by Latin subscripts
(ranging over \{1,2,3\}, unless otherwise specified). Summation over
repeated subscripts and other typical conventions for differential
operations are implied, such as a superposed dot or a comma followed by a
subscript to denote partial derivative with respect to time or the
corresponding coordinate. Occasionally, we shall use bold--face character
and typical notations for vectors and operations upon them.

We study the boundary--final value problems associated with the linear
thermoelasticity. Following \cite{[9],IesanCiarletta}, the local balance
equations become 
\begin{equation}
\begin{array}{lcr}
S_{ji,j}+\rho f_{i}=\rho \ddot{u}_{i}, & \qquad & \hbox{balance of momentum},
\\[3mm] 
h_{i,i}+g+\rho \ell =\rho \chi \ddot{\varphi},\; &  & 
\hbox{balance of
equilibrated stress}, \\[3mm] 
\rho \theta _{0}\dot{\eta}=q_{i,i}+\rho r, & \text{in\ }B\times (-\infty
,0)\smallskip & \hbox{energy equation}.
\end{array}
\label{2.3}
\end{equation}
In these equations, ${\bf u}$ is the displacement vector fields; $\varphi $
is the change in volume fraction starting from the reference configuration; $%
\theta $ is the temperature variation from the uniform reference temperature 
$\theta _{0}(>0)$. Moreover, ${\bf S}$ and ${\bf f}$ are the stress tensor
and body force, respectively; ${\bf h},g$ and $\ell $ are the equilibrated
stress vector, intrinsic and extrinsic equilibrated body force,
respectively; $\eta ,$ ${\bf q}$ and $r$ are the specific entropy, the heat
flux vector and the extrinsic heat supply, respectively. Finally, $\rho $
and $\stackrel{}{\chi }$ are the bulk mass density and equilibrated inertia
in the reference state, respectively. We assume that ${\bf f},\;\ell $ and $%
r $ are the continuous functions on $\bar{B}\times (-\infty ,0],$ with $\bar{%
B} $ closure of $B$.

By denoting ${\bf U}\equiv \{{\bf u},\varphi ,\theta \},$ it follows the
strain fields are 
\begin{equation}
e_{ij}=\frac{1}{2}(u_{i,j}+u_{j,i}),\qquad \,\gamma _{i}=\varphi
_{,i},\qquad \kappa _{i}=\theta _{,i},\qquad \qquad \hbox{in
}\bar{B}\times (-\infty ,0].  \label{2.1}
\end{equation}
The surface tractions ${\bf s},$ the surface equilibrated stress $h$ and the
boundary heat flux $q$ are 
\begin{equation}
s_{i}=S_{ji}n_{j},\qquad \qquad h=h_{j}n_{j},\qquad \qquad q=q_{j}n_{j},
\label{2.9}
\end{equation}
where ${\bf n}$ is the outward unit normal vector to the boundary surface.

We consider the porous bodies with the memory effect for the intrinsic
equilibrated body forces. Our attention is on the materials, which are
initially free from stress and the intrinsic equilibrated body force, the
entropy and the heat flux rate are equal zero. The constitutive equations
are given by 
\begin{equation}
\begin{array}{l}
S_{ij}=C_{ijrs}e_{rs}+D_{ijs}\gamma _{s}+B_{ij}\varphi -M_{ij}\theta , \\%
[3mm] 
h_{i}=D_{rsi}e_{rs}+A_{ij}\gamma _{j}+b_{i}\varphi -a_{i}\theta , \\[3mm] 
g=-\tau \dot{\varphi}+G, \\ 
G=-B_{ij}e_{ij}-b_{i}\gamma _{i}-\xi \varphi +m\theta ,\qquad \\[3mm] 
\rho \eta =M_{kl}e_{kl}+m\varphi +a_{i}\gamma _{i}+a\theta , \\[3mm] 
q_{i}=K_{ij}\kappa _{j},\quad \quad \quad \quad \quad \quad \quad \quad
\quad \quad \quad \quad \quad \quad \hbox{in }\bar{B}\times (-\infty ,0].
\end{array}
\label{2.2}
\end{equation}

The material coefficients satisfy the relations 
\begin{equation}
C_{ijrs}=C_{rsij}=C_{jirs},\quad D_{ijr}=D_{jir},\quad A_{ij}=A_{ji},\quad
B_{ij}=B_{ji},\quad M_{ij}=M_{ji},\   \label{2.4}
\end{equation}
and 
\begin{equation}
\tau \geq 0.
\end{equation}
Throughout this article, we assume the bulk mass density $\rho ,$ the
equilibrated inertia $\chi $ and the constant heat $a$ are strictly positive.

The constant conductivity tensor ${\bf K}$ is a symmetric positive definite
tensor; thus, there exist the positive constants $k_{m}$ and $k_{M}$ such
that 
\begin{equation}
k_{m}\kappa _{i}\kappa _{i}\leq K_{ij}\kappa _{i}\kappa _{j}\leq k_{M}\kappa
_{i}\kappa _{i}\qquad \qquad \forall {\bf \kappa }.\   \label{theta}
\end{equation}
The constants $k_{m}$ and $k_{M}$ are the minimum and maximum conductivity
moduli for ${\bf K.\ }$

It follows from Schwarz's inequality and the last inequality that

\begin{equation}
q_{i}q_{i}=K_{ij}\kappa _{j}q_{i}\leq (K_{ij}\kappa _{i}\kappa
_{j})^{1/2}(K_{rs}q_{r}q_{s})^{1/2}\leq (K_{ij}\kappa _{i}\kappa
_{j})^{1/2}(k_{M}q_{s}q_{s})^{1/2},
\end{equation}
so that 
\begin{equation}
q_{i}q_{i}\leq k_{M}K_{ij}\kappa _{i}\kappa _{j}.\   \label{210a}
\end{equation}

For what follows, it is useful to introduce the vector space ${\cal E}$ of
all vector fields of the form

\begin{equation}
{\bf E}\equiv \{E_{ij},\chi _{1}\pi _{i},\psi \},\qquad \qquad \text{with}%
\qquad E_{ij}=E_{ji},\qquad \text{and}\qquad \chi _{1}=\sqrt{\chi }.\ 
\label{E}
\end{equation}
Moreover, for each ${\bf E}$ $\in {\cal E}$ we define the vector field ${\bf 
\hat{S}}({\bf E})$ as 
\[
{\bf \hat{S}}({\bf E})\equiv \left\{ \hat{S}_{ji},\chi _{1}\left( \frac{1}{%
\chi }\hat{h}_{i}\right) ,-\hat{G}\right\} ,\ 
\]
where 
\begin{equation}
\begin{array}{l}
\hat{S}_{ij}=C_{ijrs}E_{rs}+D_{ijs}\pi _{s}+B_{ij}\psi ,\quad \\[3mm] 
\hat{h}_{i}=D_{rsi}E_{rs}+A_{ij}\pi _{j}+b_{i}\psi , \\[3mm] 
\hat{G}=-B_{ij}E_{ij}-b_{i}\pi _{i}-\xi \psi ,\ 
\end{array}
\label{hatS}
\end{equation}
and the coefficients obey the symmetry relations (\ref{2.4}). The vector
field ${\bf \hat{S}}({\bf E})$ belongs to ${\cal E}$, too.

Now, for any ${\bf E},{\bf \bar{E}}$ $\in {\cal E},$ we consider the
following bilinear form 
\begin{equation}
\begin{array}{l}
2{\cal F}({\bf E},{\bf \bar{E}})=C_{ijrs}E_{ij}\bar{E}_{rs}+\xi \psi \bar{%
\psi}+A_{ij}\pi _{i}\bar{\pi}_{j}+B_{ij}(E_{ij}\bar{\psi}+\bar{E}_{ij}\psi )+
\\[3mm] 
\qquad \qquad +D_{ijs}(E_{ij}\bar{\pi}_{s}+\bar{E}_{ij}\pi _{s})+b_{i}(\psi 
\bar{\pi}_{i}+\bar{\psi}\pi _{i}),
\end{array}
\label{FiEE}
\end{equation}
where ${\bf \bar{E}}\equiv \{\bar{E}_{ij},\chi _{1}^{{}}\bar{\pi}_{i},\bar{%
\psi}\}.$ From (\ref{hatS}) we have 
\begin{equation}
2{\cal F}({\bf E},{\bf \bar{E}})=[\hat{S}_{ji}\bar{E}_{ij}+\hat{h}_{i}\bar{%
\pi}_{i}-\hat{G}\bar{\psi}]\quad \text{and}\quad 2{\cal F}({\bf E},{\bf E})=[%
\hat{S}_{ji}E_{ij}+\hat{h}_{i}\pi _{i}-\hat{G}\psi ].  \label{1ener}
\end{equation}
By the constitutive equations, we prove 
\begin{equation}
{\cal F}({\bf E},{\bf \bar{E}})={\cal F}({\bf \bar{E}},{\bf E}),\qquad
\qquad \qquad \forall {\bf E,\bar{E}\in }{\cal E}.
\end{equation}
The Cauchy-Schwarz's inequality implies 
\begin{equation}
{\cal F}({\bf E},{\bf \bar{E}})\leq \lbrack \hat{W}({\bf E})]^{1/2}[\hat{W}(%
{\bf \bar{E}})]^{1/2},\qquad \qquad \forall {\bf E,\bar{E}\in }{\cal E},
\label{FSETL}
\end{equation}
where $\hat{W}$ is the quadratic form associated to ${\cal F}$ 
\begin{equation}
\begin{array}{l}
2\hat{W}({\bf E})=2{\cal F}({\bf E},{\bf E})=C_{ijrs}E_{ij}E_{rs}+\xi \psi
^{2}+A_{ij}\pi _{i}\pi _{j}+2B_{ij}\psi E_{ij}+ \\[3mm] 
\qquad \qquad \qquad \qquad \qquad +2D_{ijs}E_{ij}\pi _{s}+2b_{i}\psi \pi
_{i}.\ 
\end{array}
\end{equation}
We assume that $\hat{W}({\bf E})$ is a positive definite quadratic form and,
consequently, we have 
\begin{equation}
\mu _{m}\left( E_{ij}E_{_{ij}}+\stackrel{}{\chi }\pi _{i}\pi _{i}+\psi
^{2}\right) \leq 2\hat{W}({\bf E})\leq \mu _{M}\left( E_{ij}E_{_{ij}}+%
\stackrel{}{\chi }\pi _{i}\pi _{i}+\psi ^{2}\right) .\   \label{stel2}
\end{equation}
By setting ${\bf E=\hat{S}}({\bf E)}$ in (\ref{stel2}) and (\ref{1ener}), we
prove 
\begin{equation}
2\hat{W}({\bf \hat{S}}({\bf E)})\leq \mu _{M}\left( \hat{S}_{ji}\hat{S}_{ji}+%
\frac{1}{\chi }\hat{h}_{i}^{{}}\hat{h}_{i}^{{}}+\hat{G}^{2}\right) ,
\label{WSE}
\end{equation}
and 
\begin{equation}
\begin{array}{l}
\displaystyle\hat{S}_{ji}\hat{S}_{ji}+\frac{1}{\chi }\hat{h}_{i}\hat{h}_{i}+%
\hat{G}^{2}=2{\cal F}({\bf E},{\bf \hat{S}}({\bf E)})\leq 2[\hat{W}({\bf E}%
)]^{1/2}[\hat{W}({\bf \hat{S}}({\bf E}))]^{1/2}\leq \\[3mm] 
\displaystyle\qquad \qquad \qquad \qquad \leq 2[\hat{W}({\bf E}%
)]_{{}}^{1/2}\mu _{M}^{1/2}\left( \tilde{S}_{ji}\tilde{S}_{ji}+\frac{1}{\chi 
}\hat{h}_{i}\hat{h}_{i}+\hat{G}^{2}\right) ^{1/2};
\end{array}
\end{equation}
so that this inequality implies 
\begin{equation}
\displaystyle\hat{S}_{ji}\hat{S}_{ji}+\frac{1}{\chi }\hat{h}_{i}\hat{h}_{i}+%
\hat{G}^{2}\leq 2\mu _{M}\hat{W}({\bf E}).\   \label{ok}
\end{equation}
If$\;{\bf E}=\left\{ e_{ij},\;\chi_1\gamma _{i},\;\varphi \right\} ,\;$then
we introduce the following notations 
\begin{equation}
\begin{array}{l}
2W^{\ast }=2\hat{W}({\bf E})=C_{ijrs}e_{ij}e_{rs}+\xi \varphi
^{2}+A_{ij}\gamma _{i}\gamma _{j}+2B_{ij}\varphi e_{ij}+2D_{ijs}e_{ij}\gamma
_{s}+2b_{i}\varphi \gamma _{i}, \\[3mm] 
\tilde{S}_{ij}=C_{ijrs}e_{rs}+D_{ijs}\gamma _{s}+B_{ij}\varphi ,\quad \quad
\quad \quad \quad \tilde{h}_{i}=D_{rsi}e_{rs}+A_{ij}\gamma _{j}+b_{i}\varphi
, \\[3mm] 
\tilde{G}=-B_{ij}e_{ij}-b_{i}\gamma _{i}-\xi \varphi .\ 
\end{array}
\label{tutto}
\end{equation}
By eqs. (\ref{hatS}), (\ref{1ener}) and (\ref{tutto}), we can see that 
\begin{equation}
2W^{\ast }=[\tilde{S}_{ji}e_{ij}+\tilde{h}_{i}\gamma _{i}-\tilde{G}\varphi
],\qquad \qquad \dot{W}^{\ast }=[\tilde{S}_{ji}\dot{e}_{ij}+\tilde{h}_{i}%
\dot{\gamma}_{i}-\tilde{G}\dot{\varphi}].\   \label{wdot}
\end{equation}
For any positive number $\epsilon $ and each second-order tensors, ${\bf L}$
and ${\bf F,}$ we have the inequality 
\begin{equation}
(L_{ij}+F_{ij})(L_{ij}+F_{ij})\leq (1+\epsilon )L_{ij}L_{ij}+(1+\frac{1}{%
\epsilon })F_{ij}F_{ij}.\   \label{tensor}
\end{equation}
By aid of eqs. (\ref{2.2}), (\ref{tutto}) and (\ref{tensor}), we obtain 
\begin{equation}
\begin{array}{l}
S_{ij}=\tilde{S}_{ij}-M_{ij}\theta ,\qquad h_{i}=\tilde{h}_{i}-a_{i}\theta
,\qquad G=\tilde{G}+m\theta ,
\end{array}
\end{equation}
and 
\begin{equation}
\begin{array}{l}
\displaystyle S_{ij}S_{ij}+\frac{1}{\stackrel{}{\chi }}h_{i}h_{i}=(\tilde{S}%
_{ij}-M_{ij}\theta )(\tilde{S}_{ij}-M_{ij}\theta )+\frac{1}{\stackrel{}{\chi 
}}(\tilde{h}_{i}-a_{i}\theta )(\tilde{h}_{i}-a_{i}\theta )\leq \\[3mm] 
\displaystyle\qquad (1+\epsilon )\tilde{S}_{ij}\tilde{S}_{ij}+(1+\frac{1}{%
\epsilon })M_{ij}M_{ij}\theta ^{2}+\frac{1}{\stackrel{}{\chi }}[(1+\epsilon )%
\tilde{h}_{i}\tilde{h}_{i}+(1+\frac{1}{\epsilon })a_{i}a_{i}\theta ^{2}]\leq
\\[3mm] 
\qquad \leq (1+\epsilon )2\mu _{M}W^{\ast }+(1+\frac{1}{\epsilon }%
)M^{2}\theta ^{2},\qquad \qquad \qquad \qquad \qquad \qquad \qquad \forall
\epsilon >0,
\end{array}
\label{2.10b}
\end{equation}
in which 
\begin{equation}
M^{2}=\max_{\bar{B}}(M_{ij}M_{ij}+\frac{1}{\stackrel{}{\chi }}a_{i}a_{i}).\ 
\end{equation}

We consider the boundary-final value problem ${\cal P}$ defined by the
equations of motion (\ref{2.3}), the geometrical equations (\ref{2.1}) and
the constitutive equations (\ref{2.2}) and the following final-boundary
conditions 
\begin{equation}
\begin{array}{l}
u_{i}({\bf x},0)=u_{i}^{0}({\bf x)},\quad \quad \dot{u}_{i}({\bf x},0)=\dot{u%
}_{i}^{0}({\bf x)},\quad \quad \varphi ({\bf x},0{\bf )}=\varphi ^{0}({\bf %
x),} \\[3mm] 
\dot{\varphi}({\bf x},0{\bf )}=\dot{\varphi}^{0}({\bf x)},\quad \quad \theta
({\bf x},0{\bf )}=\theta ^{0}({\bf x)},\qquad \qquad {\bf x}\in {B},\ 
\end{array}
\label{inizio}
\end{equation}
and 
\begin{equation}
\begin{array}{l}
u_{i}=u_{i}^{\ast }\quad \hbox{ on }\bar{\Sigma}_{1}\times (-\infty
,0],\qquad s_{i}=s_{i}^{\ast }\quad \hbox{
on }\Sigma _{2}\times (-\infty ,0], \\[3mm] 
\varphi =\varphi ^{\ast }\quad \hbox{ on }\bar{\Sigma}_{3}\times (-\infty
,0],\qquad h=h^{\ast }\quad \hbox{ on }\Sigma _{4}\times (-\infty ,0], \\%
[3mm] 
\theta =\theta ^{\ast }\quad \hbox{ on }\bar{\Sigma}_{5}\times (-\infty
,0],\qquad q=q^{\ast }\quad \hbox{ on }\Sigma _{6}\times (-\infty ,0],
\end{array}
\label{bordo}
\end{equation}
where $\Sigma _{i}$ ($i=1,...\ ,6$) are the subsets of $\partial B$ such
that 
\[
\bar{\Sigma}_{1}\cup \Sigma _{2}=\bar{\Sigma}_{3}\cup \Sigma _{4}=\bar{\Sigma%
}_{5}\cup \Sigma _{6}=\partial B,\qquad \Sigma _{1}\cap \Sigma _{2}=\Sigma
_{3}\cap \Sigma _{4}=\Sigma _{5}\cap \Sigma _{6}=\emptyset .\ 
\]
The terms on the right-hand of eqs. (\ref{inizio}) and (\ref{bordo}) are
prescribed continuous functions.

For further convenience we use an appropriate change of time variable and
notations to transform the boundary-final value problem ${\cal P}$ in the
boundary-initial value problem ${\cal P}^{{\cal \ast }}$ defined by the
following equations 
\begin{equation}
\begin{array}{lc}
S_{ji,j}+\rho f_{i}=\rho \ddot{u}_{i}, & \qquad \\[3mm] 
h_{i,i}+g+\rho \ell =\rho \chi \ddot{\varphi},\; &  \\[3mm] 
-\rho \theta _{0}\dot{\eta}=q_{i,i}+\rho r, & \text{in\ }B\times (0,+\infty
),\smallskip
\end{array}
\label{2.3*}
\end{equation}
\begin{equation}
e_{ij}=\frac{1}{2}(u_{i,j}+u_{j,i}),\qquad \,\gamma _{i}=\varphi
_{,i},\qquad \kappa _{i}=\theta _{,i},\qquad \qquad \hbox{in }\bar{B}\times
\lbrack 0,+\infty ),
\end{equation}
and 
\begin{equation}
\begin{array}{l}
S_{ij}=C_{ijrs}e_{rs}+D_{ijs}\gamma _{s}+B_{ij}\varphi -M_{ij}\theta , \\%
[3mm] 
h_{i}=D_{rsi}e_{rs}+A_{ij}\gamma _{j}+b_{i}\varphi -a_{i}\theta , \\[3mm] 
g=\tau \dot{\varphi}+G, \\ 
G=-B_{ij}e_{ij}-b_{i}\gamma _{i}-\xi \varphi +m\theta ,\qquad \\[3mm] 
\rho \eta =M_{kl}e_{kl}+a_{i}\gamma _{i}+m\varphi +a\theta , \\[3mm] 
q_{i}=K_{ij}\kappa _{j},\quad \quad \quad \quad \quad \quad \quad \quad
\quad \quad \quad \quad \quad \quad \hbox{in }\bar{B}\times \lbrack
0,+\infty ),
\end{array}
\label{2.2*}
\end{equation}
with the initial conditions 
\begin{equation}
\begin{array}{l}
u_{i}({\bf x},0)=u_{i}^{0}({\bf x)},\quad \quad \dot{u}_{i}({\bf x},0)=\dot{u%
}_{i}^{0}({\bf x)},\quad \quad \varphi ({\bf x},0{\bf )}=\varphi ^{0}({\bf %
x),} \\[3mm] 
\dot{\varphi}({\bf x},0{\bf )}=\dot{\varphi}^{0}({\bf x)},\quad \quad \theta
({\bf x},0{\bf )}=\theta ^{0}({\bf x)},\qquad \qquad {\bf x}\in {B},\ 
\end{array}
\label{inizio*}
\end{equation}
and the boundary conditions 
\begin{equation}
\begin{array}{l}
u_{i}=u_{i}^{\ast }\quad \hbox{ on }\bar{\Sigma}_{1}\times \lbrack 0,+\infty
),\qquad s_{i}=s_{i}^{\ast }\quad \hbox{
on }\Sigma _{2}\times \lbrack 0,+\infty ), \\[3mm] 
\varphi =\varphi ^{\ast }\quad \hbox{ on }\bar{\Sigma}_{3}\times \lbrack
0,+\infty ),\qquad h=h^{\ast }\quad \hbox{ on }\Sigma _{4}\times \lbrack
0,+\infty ), \\[3mm] 
\theta =\theta ^{\ast }\quad \hbox{ on }\bar{\Sigma}_{5}\times \lbrack
0,+\infty ),\qquad q=q^{\ast }\quad \hbox{ on }\Sigma _{6}\times \lbrack
0,+\infty ).\ 
\end{array}
\label{bordo*}
\end{equation}

We define as solution of the boundary-initial value problem ${\cal P}^{{\cal %
\ast }}$ a process ${\bf \pi }=\{{\bf u}${\bf , }${\bf e,\ S,\ }\varphi {\bf %
,\ \gamma ,\ h,\ }g{\bf ,\ }\theta {\bf ,\ \kappa ,\ q,\ }\eta \}$ that
satisfies eqs. (\ref{2.3*})--(\ref{bordo*})\ and

i. \thinspace \thinspace \thinspace $u_{i},\varphi \in C^{2,2}(\bar{B}\times
\lbrack 0,+\infty )),\;\theta \in C^{1,1}(\bar{B}\times \lbrack 0,+\infty ))$%
;

ii. \thinspace \thinspace\ $e_{ij}=e_{ji},\;\gamma _{i},\kappa _{i}\in
C^{1,1}(\bar{B}\times \lbrack 0,+\infty ))$;

iii. \thinspace\ $S_{ij}=S_{ji},\;h_{i},q_{i},g\in C^{1,0}(\bar{B}\times
\lbrack 0,+\infty ))$; $\eta \in C^{0,1}(\bar{B}\times \lbrack 0,+\infty
)).\ $

\section{Saint-Venant's Principle}

Let us consider a given time $T$ $\in (0,+\infty )$ and a given ( external )
data ${\cal D}=\{{\bf f}${\bf , }$\ell $, $r$; $u_{i}^{0}${\bf , }$\dot{u}%
_{i}^{0}$, $\varphi ^{0}$, $\dot{\varphi}^{0}$, $\theta ^{0}$; $u_{i}^{\ast
} ${\bf , }$s_{i}^{\ast }$, $\varphi ^{\ast }$, $h^{\ast }$, $\theta ^{\ast
} $, $q^{\ast }\}$ in the problem ${\cal P}^{{\cal \ast }}$. We denote by $%
\widehat{D}_{T}$ \ the support of the initial and boundary data, the body
force, the extrinsic equilibrated body force and the heat supply on the time
interval $[0,T]$, i. e. the set of all ${\bf x}{\in }\bar{B}$ such that: 
\newline
i. \thinspace \thinspace \thinspace\ if ${\bf x}{\in B},$ then 
\begin{equation}
\begin{array}{l}
\;\;u_{i}^{0}({\bf x})\neq 0\;\hbox{ or }\;\dot{u}_{i}^{0}({\bf x})\ \neq \
0,\;\hbox{ or }\;\varphi ^{0}({\bf x})\neq 0,\;\hbox{ or }\;\dot{\varphi}%
^{0}({\bf x})\neq \ 0,\;\hbox{ or }\;\theta ^{0}({\bf x})\neq 0, \\[3mm] 
\hbox{ or }\;f_{i}({\bf x},s)\neq 0,\;\hbox{ or }\;\ell ({\bf x},s)\neq 0,\;%
\hbox{ or }\;r({\bf x},s)\neq 0\;\hbox{ for some }\;s\in \lbrack 0,T];
\end{array}
\label{2.11}
\end{equation}
ii. if ${\bf x}\in \partial B,$ then 
\begin{equation}
\begin{array}{l}
\hbox{    }\;\;\;u_{i}^{\ast }({\bf x},s)\neq 0\;\hbox{ for some }({\bf x}%
,s)\in \bar{\Sigma}_{1}\times \lbrack 0,T],\; \\[3mm] 
\hbox{
or }s_{i}^{\ast }({\bf x},s)\neq 0\;\hbox{ for some }({\bf x},s)\in \Sigma
_{2}\times \lbrack 0,T],\; \\[3mm] 
\hbox{
or }\varphi ^{\ast }({\bf x},s)\neq 0\;\hbox{ for some }({\bf x},s)\in \bar{%
\Sigma}_{3}\times \lbrack 0,T],\; \\[3mm] 
\hbox{
or }h^{\ast }({\bf x},s)\neq 0\;\hbox{ for some }({\bf x},s)\in \Sigma
_{4}\times \lbrack 0,T],\; \\[3mm] 
\hbox{
or }\theta ^{\ast }({\bf x},s)\neq 0\;\hbox{ for some }({\bf x},s)\in \bar{%
\Sigma}_{5}\times \lbrack 0,T],\; \\[3mm] 
\hbox{
or }q^{\ast }({\bf x},s)\neq 0\;\hbox{ for some }({\bf x},s)\in \Sigma
_{6}\times \lbrack 0,T].\ 
\end{array}
\label{2.13}
\end{equation}
We assume that $\widehat{D}_{T}$ is a bounded set. We consider a non-empty
bounded regular region $\widehat{D}_{T}^{\ast }$ such that ${\widehat{D}_{T}}%
\subset {\widehat{D}_{T}^{\ast }}\subset \bar{B}$. We note that

i. \thinspace \thinspace \thinspace\ if $\emptyset \neq {\widehat{D}_{T}}$,
then we choose $\widehat{D}_{T}^{\ast }$ to be the smallest bounded regular
region in $\bar{B}$ that includes $\widehat{D}_{T}$ ; in particular, we set $%
{\widehat{D}_{T}^{\ast }}={\widehat{D}_{T}}$ if $\widehat{D}_{T}$ is also a
regular region;

ii. \thinspace \thinspace\ if ${\widehat{D}_{T}}=\emptyset $, then $\widehat{%
D}_{T}^{\ast }$ may be chosen in an arbitrary way.

Now, we mean the set $D_{r},$ by 
\begin{equation}
D_{r}=\{{\bf x}{\in }\bar{B}:{\widehat{D}_{T}^{\ast }}\cap {\bar{\Sigma}(%
{\bf x,r)}}\neq \emptyset \},\qquad r\geq 0,  \label{2.14}
\end{equation}
where $\bar{\Sigma}({\bf x},r)$ is the closed ball with radius $r$ and
center at ${\bf x}$. Clearly, ${\widehat{D}_{T}}\subseteq {\widehat{D}%
_{T}^{\ast }=D}_{{0}}\subset D_{r}$ $(r>0).$ Further, we set $%
B_{r}=B\setminus D_{r};$ we have $B_{r_{2}}\supset B_{r_{1}},$ and $%
B(r_{1},r_{2})=B_{r_{2}}\setminus B_{r_{1}}$ for $r_{1}>r_{2}$. Let $L$ be
the diameter of $B_{0}.$ The surface $S_{r}$ is the subsurface of $\partial
B_{r}$ contained inside $B$ and whose outward unit normal vector is oriented
to the exterior of $D_{r}.\ $

In what follows we need the below lemma

{\bf Lemma 1. }Let ${\bf \pi }=\left\{ {\bf u,e,S},\varphi ,{\bf \gamma ,h,}%
g,\theta ,{\bf \kappa ,q},\eta \right\} $ be a the solution of ${\cal P}^{%
{\cal \ast }}.$ Then, for every regular region $P\subseteq B$ with regular
boundary $\partial P$ and for each $t\in \lbrack 0,T]$, we obtain

\begin{equation}
\begin{array}{l}
\displaystyle\int_{0}^{t}\int_{P}e^{\lambda s}\{\frac{\lambda }{2}[\rho \dot{%
u}_{i}(s)\dot{u}_{i}(s)+\rho \chi \dot{\varphi}^{2}(s)+a\theta
^{2}(s)+2W^{\ast }(s)]+\tau \dot{\varphi}^{2}(s)+ \\[3mm] 
\displaystyle\qquad +\frac{1}{\theta _{0}}K_{ij}\kappa _{i}(s)\kappa
_{j}(s)\}dvds=\int_{P}e^{\lambda t}\frac{1}{2}[\rho \dot{u}_{i}(t)\dot{u}%
_{i}(t)+\rho \chi \dot{\varphi}^{2}(t)+a\theta ^{2}(t)+ \\[3mm] 
\displaystyle\qquad +2W^{\ast }(t)]dv-\int_{0}^{t}\int_{\partial
P}e^{\lambda s}\left[ s_{i}(s)\dot{u}_{i}(s)+h(s)\dot{\varphi}(s)-\frac{q}{%
\theta _{0}}\theta (s)\right] dads+ \\[3mm] 
\displaystyle\qquad -\int_{0}^{t}\int_{P}e^{\lambda s}\left[ \rho f_{i}(s)%
\dot{u}_{i}(s)+\rho \ell \dot{\varphi}(s)-\frac{\rho r}{\theta _{0}}\theta
(s)\right] dvds+ \\ 
\displaystyle\qquad -\int_{P}\frac{1}{2}[\rho \dot{u}_{i}^{0}\dot{u}%
_{i}^{0}+\rho \chi \dot{\varphi}^{0}\dot{\varphi}^{0}+a\theta ^{0}\theta
^{0}+2W^{\ast }(0)]dv,
\end{array}
\end{equation}
where $\lambda $ is a prescribed positive parameter.

{\bf Proof.} By eqs. (\ref{2.3*})--(\ref{2.2*}) we deduce that 
\begin{equation}
\begin{array}{l}
\displaystyle\frac{\partial }{\partial s}\left\{ \frac{1}{2}[\rho \dot{u}%
_{i}(s)\dot{u}_{i}(s)+\rho \chi \dot{\varphi}^{2}(s)+a\theta
^{2}(s)+2W^{\ast }(s)]\right\} =\tau \dot{\varphi}^{2}(s)+ \\[3mm] 
\displaystyle\qquad +\frac{1}{\theta _{0}}K_{ij}\kappa _{i}(s)\kappa
_{j}(s)+\rho f_{i}(s)\dot{u}_{i}(s)+\rho \ell \dot{\varphi}(s)-\frac{\rho r}{%
\theta _{0}}\theta (s)+ \\[3mm] 
\displaystyle\qquad +\left[ S_{ji}(s)\dot{u}_{i}(s)+h_{j}(s)\dot{\varphi}(s)-%
\frac{q_{j}}{\theta _{0}}\theta (s)\right] _{,j}.\ 
\end{array}
\end{equation}

Thus, we can see that 
\begin{equation}
\begin{array}{l}
\displaystyle\frac{\partial }{\partial s}\left\{ e^{\lambda s}\frac{1}{2}[%
\rho \dot{u}_{i}(s)\dot{u}_{i}(s)+\rho \chi \dot{\varphi}^{2}(s)+a\theta
^{2}(s)+2W^{\ast }(s)]\right\} = \\[3mm] 
\displaystyle\qquad =e^{\lambda s}\frac{\lambda }{2}[\rho \dot{u}_{i}(s)\dot{%
u}_{i}(s)+\rho \chi \dot{\varphi}^{2}(s)+a\theta ^{2}(s)+2W^{\ast }(s)%
]+e^{\lambda s}\tau \dot{\varphi}^{2}(s)+ \\[3mm] 
\displaystyle\qquad +e^{\lambda s}\frac{1}{\theta _{0}}K_{ij}\kappa
_{i}(s)\kappa _{j}(s)+e^{\lambda s}\left[ \rho f_{i}(s)\dot{u}_{i}(s)+\rho
\ell \dot{\varphi}(s)-\frac{\rho r}{\theta _{0}}\theta (s)\right] + \\[3mm] 
\displaystyle\qquad +e^{\lambda s}\left[ S_{ji}(s)\dot{u}_{i}(s)+h_{j}(s)%
\dot{\varphi}(s)-\frac{q_{j}}{\theta _{0}}\theta (s)\right] _{,j}.\ 
\end{array}
\end{equation}
If we integrate this relation over $P\times \lbrack 0,T],$ then we obtain
the desired result with the help of the divergence theorem and eqs. (\ref
{2.9}). $\bullet $\bigskip

For a prescribed strictly positive parameter $\lambda $ and for any $r\in
\lbrack 0,L],$ $t\in \lbrack 0,T]$, we associate with the solution ${\bf \pi 
}$ the following time--weighted volume measure ${\cal E}(r,t)$ $(>0)$

\begin{equation}
\begin{array}{l}
\displaystyle{\cal E}(r,t)=\int_{0}^{t}\int_{B_{r}}e^{\lambda s}\{\frac{%
\lambda }{2}[\rho \dot{u}_{i}(s)\dot{u}_{i}(s)+\rho \chi \dot{\varphi}%
^{2}(s)+a\theta ^{2}(s)+2W^{\ast }(s)]+ \\[3mm] 
\displaystyle\qquad \qquad +\tau \dot{\varphi}^{2}(s)+\frac{1}{\theta _{0}}%
K_{ij}\kappa _{i}(s)\kappa _{j}(s)\}dvds.
\end{array}
\end{equation}
The parameter $\lambda ,$ in the above function, is characteristic for the
considered measure.

Taking into account that, for $r_{1}\geq r_{2}$ 
\begin{equation}
\begin{array}{l}
\displaystyle{\cal E}(r_{1},t)-{\cal E}(r_{2},t)=-\int_{0}^{t}%
\int_{B(r_{1},r_{2})}e^{\lambda s}\{\frac{\lambda }{2}[\rho \dot{u}_{i}(s)%
\dot{u}_{i}(s)+\rho \chi \dot{\varphi}^{2}(s)+ \\[3mm] 
\displaystyle\qquad +a\theta ^{2}(s)+2W^{\ast }(s)]+\tau \dot{\varphi}%
^{2}(s)+\frac{1}{\theta _{0}}K_{ij}\kappa _{i}(s)\kappa _{j}(s)\}dvds,
\end{array}
\label{2.25}
\end{equation}
it is a simple matter to prove the following lemma.

{\bf Lemma 2.} Let ${\bf \pi }$\ be a solution of initial-boundary-value
problem ${\cal P}^{{\cal \ast }}$\ and $\widehat{D}_{T}$\ be the bounded
support of the external data ${\cal D}$ on the time interval $[0,T]$. Then,
the corresponding time--weighted volume measure satisfies the following
properties

(i) ${\cal E}(r,t)$\ is a non--increasing function with respect to $r,$\ i.
e. 
\begin{equation}
{\cal E}(r_{1},t)\leq {\cal E}(r_{2},t),\qquad \text{with }r_{1}\geq
r_{2},t\in \lbrack 0,T].\   \label{positiva}
\end{equation}

(ii) ${\cal E}(r,t)$\ is a continuous differentiable function on $r\in
\lbrack 0,L]$, $t\in \lbrack 0,T]$\ and 
\begin{equation}
\begin{array}{l}
\displaystyle\frac{\partial }{\partial r}{\cal E}(r,t)=-\int_{0}^{t}%
\int_{S_{r}}e^{\lambda s}\{\frac{\lambda }{2}[\rho \dot{u}_{i}(s)\dot{u}%
_{i}(s)+\rho \chi \dot{\varphi}^{2}(s)+a\theta ^{2}(s)+2W^{\ast }(s)]+ \\%
[3mm] 
\displaystyle\qquad \qquad +\tau \dot{\varphi}^{2}(s)+\frac{1}{\theta _{0}}%
K_{ij}\kappa _{i}(s)\kappa _{j}(s)\}dads,
\end{array}
\label{derispaE}
\end{equation}
\begin{equation}
\begin{array}{l}
\displaystyle\frac{\partial }{\partial t}{\cal E}(r,t)=\int_{B_{r}}e^{%
\lambda t}\{\frac{\lambda }{2}[\rho \dot{u}_{i}(t)\dot{u}_{i}(t)+\rho \chi 
\dot{\varphi}^{2}(t)+a\theta ^{2}(t)+2W^{\ast }(t)]+ \\[3mm] 
\displaystyle\qquad +\tau \dot{\varphi}^{2}(t)+\frac{1}{\theta _{0}}%
K_{ij}\kappa _{i}(t)\kappa _{j}(t)\}dv.\ 
\end{array}
\label{deritemE}
\end{equation}

{\bf Lemma 3.} Let ${\bf \pi }$ be a solution of the initial-boundary-value
problem ${\cal P}^{{\cal \ast }}$\ and $\widehat{D}_{T}$\ be the bounded
support of the external data ${\cal D}$ on the time interval $[0,T]$. Then, $%
{\cal E}(r,t)$\ satisfies the following first--order differential inequality 
\begin{equation}
{\cal E}(r,t)\leq -\ \frac{\zeta }{\lambda }\frac{\partial }{\partial r}%
{\cal E}(r,t)+\ \frac{1}{\lambda }\frac{\partial }{\partial t}{\cal E}%
(r,t),\qquad \qquad \forall r\in \lbrack 0,L],t\in \lbrack 0,T],
\label{2.20}
\end{equation}
where 
\begin{equation}
\zeta (\lambda )=\displaystyle\sqrt{\frac{\mu _{M}}{\rho }(1+\varepsilon )},
\label{2.21}
\end{equation}
and 
\begin{equation}
1+\varepsilon =\frac{1}{2}+\frac{M^{2}}{2a\rho \mu _{M}}+\frac{\lambda k_{M}%
}{4\theta _{0}a\mu _{M}}+\sqrt{\left( \frac{1}{2}-\frac{M^{2}}{2a\rho \mu
_{M}}-\frac{\lambda k_{M}}{4a\theta _{0}\mu _{M}}\right) ^{2}+\frac{M^{2}}{%
a\rho \mu _{M}}\ .}  \label{algebrica}
\end{equation}

{\bf Proof.} Taking $P=B_{r}$ into Lemma 1, we have 
\begin{equation}
\begin{array}{l}
\displaystyle{\cal E}(r,t)=\int_{B_{r}}e^{\lambda t}\frac{1}{2}[\rho \dot{u}%
_{i}(t)\dot{u}_{i}(t)+\rho \chi \dot{\varphi}^{2}(t)+a\theta
^{2}(t)+2W^{\ast }(t)]dv- \\[3mm] 
\displaystyle-\int_{0}^{t}\int_{S_{r}}e^{\lambda s}\left[ s_{i}(t)\dot{u}%
_{i}(t)+h(t)\dot{\varphi}(t)-\frac{q}{\theta _{0}}\theta (t)\right] dadt.\ 
\end{array}
\label{Eme}
\end{equation}
It follows from (\ref{deritemE}) and (\ref{Eme}) 
\begin{equation}
\displaystyle\int_{B_{r}}e^{\lambda t}\frac{1}{2}[\rho \dot{u}_{i}(t)\dot{u}%
_{i}(t)+\rho \chi \dot{\varphi}^{2}(t)+a\theta ^{2}(t)+2W^{\ast }(t)]dv\leq 
\frac{1}{\lambda }\frac{\partial {\cal E}}{\partial t}(r,t).\ 
\label{mideri}
\end{equation}
If we use Schwarz's inequality and the arithmetic--geometric mean
inequality, we obtain

\begin{equation}
\begin{array}{l}
\displaystyle|S_{ji}(s)n_{j}\dot{u}_{i}(s)+h_{j}(s)n_{j}\dot{\varphi}(s)-%
\frac{1}{\theta _{0}}\theta (s)q_{j}(s)n_{j}|\leq \\[3mm] 
\displaystyle\ \ \leq \frac{1}{\lambda \varepsilon _{1}}[\frac{\lambda }{2}%
\rho \dot{u}_{i}(s)\dot{u}_{i}(s)]+\frac{\varepsilon _{1}}{\lambda \rho }%
\left[ \frac{\lambda }{2}S_{ji}(s)S_{ji}(s)\right] + \\[3mm] 
\displaystyle\ \ +\frac{1}{\lambda \varepsilon _{1}}[\frac{\lambda }{2}(\rho
\chi +\frac{2\tau }{\lambda })\dot{\varphi}^{2}(s)]+\frac{\varepsilon _{1}}{%
\displaystyle\lambda (\rho +\frac{2\tau }{\lambda \chi })}\left[ \frac{%
\displaystyle\lambda h_{j}(s)h_{j}(s)}{\displaystyle2\chi }\right] + \\[3mm] 
\displaystyle\ \ +\frac{1}{\theta _{0}\lambda \varepsilon _{2}}[\frac{%
\lambda }{2}a\theta ^{2}(s)]+\frac{\varepsilon _{2}}{2a}[\frac{1}{\theta _{0}%
}q_{j}(s)q_{j}(s)],\qquad \forall \varepsilon _{1},\varepsilon _{2}>0.\ 
\end{array}
\label{49}
\end{equation}
By eqs. (\ref{210a}), (\ref{2.10b}) and (\ref{49}), we deduce that 
\begin{equation}
\begin{array}{l}
\displaystyle|S_{ji}(s)n_{j}\dot{u}_{i}(s)+h_{j}(s)n_{j}\dot{\varphi}(s)-%
\frac{1}{\theta _{0}}\theta (s)q_{j}(s)n_{j}|\leq \\[3mm] 
\displaystyle\ \ \leq \frac{1}{\lambda \varepsilon _{1}}[\frac{\lambda }{2}%
(\rho \dot{u}_{i}(s)\dot{u}_{i}(s)+\rho \chi \dot{\varphi}^{2}(s))+\tau \dot{%
\varphi}^{2}(s)]+\frac{\varepsilon _{1}(1+\epsilon )\mu _{M}}{\lambda \rho }%
\left[ \lambda W^{\ast }(s)\right] + \\[3mm] 
\displaystyle\ \ +\left[ \frac{\varepsilon _{1}M^{2}}{\lambda \rho a}(1+%
\frac{1}{\epsilon })+\frac{1}{\lambda \theta _{0}\varepsilon _{2}}\right] [%
\frac{\lambda }{2}a\theta ^{2}(s)]+\frac{\varepsilon _{2}k_{M}}{2a}[\frac{1}{%
\theta _{0}}K_{ij}\kappa _{i}(s)\kappa _{j}(s)],\qquad \forall
\varepsilon,\varepsilon _{1},\varepsilon _{2} >0.\ 
\end{array}
\label{disug}
\end{equation}
If we consider $\varepsilon ,\varepsilon _{1},\varepsilon _{2}$ such that 
\begin{equation}
\displaystyle\frac{1}{\lambda \varepsilon _{1}}=\frac{\varepsilon
_{1}(1+\epsilon )\mu _{M}}{\lambda \rho }=\frac{\varepsilon _{1}M^{2}}{%
\lambda \rho a}(1+\frac{1}{\epsilon })+\frac{1}{\lambda \theta
_{0}\varepsilon _{2}}=\frac{\varepsilon _{2}k_{M}}{2a},\ 
\end{equation}
then 
\begin{equation}
\displaystyle\varepsilon _{1}=\frac{1}{\zeta (\lambda )},\qquad \varepsilon
_{2}=\frac{2a\zeta (\lambda )}{\lambda k_{M}},  \label{epsilons}
\end{equation}
and $\varepsilon $ satisfies to (\ref{algebrica}), i. e. $\varepsilon $ is
the positive root of the algebraic equation

\begin{equation}
\displaystyle\varepsilon ^{2}+2\varepsilon (\frac{1}{2}-\frac{M^{2}}{2a\rho
\mu _{M}}-\frac{\lambda k_{M}}{4\theta _{0}a\mu _{M}})-\frac{M^{2}}{a\rho
\mu _{M}}=0.\ 
\end{equation}
By eqs. (\ref{disug}), (\ref{epsilons}) we get for any $r\in \lbrack 0,L]$
and $t\in \lbrack 0,T]$%
\begin{equation}
\begin{array}{l}
\displaystyle-\int_{0}^{t}\int_{S_{r}}e^{\lambda s}[s_{i}(s)\dot{u}%
_{i}(s)+h_{i}(s)n_{j}\dot{\varphi}(s)-\frac{1}{\theta _{0}}\theta
(s)q_{j}(s)n_{j}]dads\leq \\[3mm] 
\leq \displaystyle|\int_{0}^{t}\int_{S_{r}}e^{\lambda s}[s_{i}(s)\dot{u}%
_{i}(s)+h_{i}(s)n_{j}\dot{\varphi}(s)-\frac{1}{\theta _{0}}\theta
(s)q_{j}(s)n_{j}]dads|\leq -\frac{\zeta }{\lambda }\frac{\partial {\cal E}}{%
\partial r}(r,t).
\end{array}
\label{mideri+}
\end{equation}

Therefore, the relations (\ref{Eme}), (\ref{mideri}), (\ref{mideri+}) lead
to (\ref{2.20}). Thus, the proof is complete. $\bullet $\bigskip

{\bf Theorem 1.} Let ${\bf \pi }$ be a solution of initial-boundary-value
problem ${\cal P}^{{\cal \ast }}$\ and $\widehat{D}_{T}$\ be the bounded
support of the external data ${\cal D}$ on the time interval $[0,T]$. For $%
\lambda $ (sufficiently large), $t_{0}\in \lbrack 0,T]$, $r_{0}\in \lbrack
0,L]$ such that 
\begin{equation}
L\leq \zeta (\lambda )t_{0}+r_{0}\leq \zeta (\lambda )T,\   \label{disequa}
\end{equation}
we have

\begin{equation}
{\cal E}(r,t_{0}+\frac{r_{0}-r}{\zeta (\lambda )})\leq {\cal E}(0,t_{0}+%
\frac{r_{0}}{\zeta (\lambda )})\,{exp}\,(-\displaystyle\frac{\lambda }{\zeta
(\lambda )}r),\qquad \qquad \hbox{ for }r\in \lbrack 0,L].\   \label{2. 38}
\end{equation}

{\bf Proof. }We put 
\begin{equation}
\displaystyle{\cal I}(r,t)=\exp (\frac{\lambda }{\zeta (\lambda )}r){\cal E}%
(r,t),\qquad \qquad \forall r\in \lbrack 0,L],{\bf \ }t\in \lbrack 0,T];
\label{I+}
\end{equation}
the inequality (\ref{2.20}) takes the following form 
\begin{equation}
\displaystyle\frac{\partial {\cal I}}{\partial r}(r,t)-\frac{1}{\zeta
(\lambda )}\frac{\partial {\cal I}}{\partial t}(r,t)\leq 0,\qquad \qquad
\forall r\in \lbrack 0,L],\;t\in \lbrack 0,T].\   \label{IT}
\end{equation}

From eqs. (\ref{2.21}) and (\ref{algebrica}) it is trivial to see $\zeta
(\lambda )\sim \lambda ^{1/2}$ for $\lambda\rightarrow\infty$ and so $\zeta
(\lambda )$ is an increasing function for sufficiently large values of $%
\lambda .$ Thus, we can choose $t_{0}$ and $r_{0}$ satisfying (\ref{disequa}%
) and hence 
\begin{equation}
\displaystyle0\leq t_{0}+\frac{r_{0}-r}{\zeta (\lambda )}\leq T\qquad \qquad
\forall r\in \lbrack 0,L].\ 
\end{equation}
By setting $\displaystyle t=t_{0}+\frac{r_{0}-r}{\zeta (\lambda )}$ in (\ref
{IT}), it follows 
\begin{equation}
\displaystyle\frac{d}{dr}[{\cal I}(r,t_{0}+\frac{r_{0}-r}{\zeta (\lambda )})%
]\leq 0,\qquad \qquad \forall r\in \lbrack 0,L],
\end{equation}
and therefore, 
\begin{equation}
0\leq {\cal I}(r,t_{0}+\frac{r_{0}-r}{\zeta (\lambda )})\leq {\cal I}%
(0,t_{0}+\frac{r_{0}}{\zeta (\lambda )}),\qquad \qquad \forall r\in \lbrack
0,L].\   \label{I<}
\end{equation}
The relations (\ref{I+}) and (\ref{I<}) lead to (\ref{2. 38}) and the proof
is complete. $\bullet $\bigskip

%\section{BIBLIOGRAFIA}

\end{document}